\documentclass[aps,pra,twocolumn,showpacs,floatfix]{revtex4}
\usepackage{epsfig}
\usepackage{graphicx}
\usepackage{dcolumn}
\usepackage{amsthm,amsmath}
\usepackage{color}

\begin{document}

\title{Relativistic equation-of-motion coupled-cluster method for the double ionization potentials
of the closed-shell atoms} 

\author{Himadri Pathak\footnote{h.pathak@ncl.res.in}$^1$, Aryya Ghosh$^1$, B. K. Sahoo$^2$,
B. P. Das$^3$, Nayana Vaval$^1$ and Sourav Pal$^1$}

\affiliation{$^1$Electronic Structure Theory Group, Physical Chemistry Division,
CSIR-National Chemical Laboratory, Pune, 411008, India}

\affiliation{$^2$Theoretical Physics Division, Physical Research Laboratory,
Ahmedabad, 380009, India}

\affiliation{$^3$Theoretical Physics and Astrophysics Group, Indian Institute of Astrophysics,
Bangalore, 560034, India}

\begin{abstract}
We report the implementation of the relativistic equation-of-motion coupled-cluster
method to calculate double ionization  spectra (DI-EOMCC) of the closed-shell atomic
systems. This method is employed to calculate the principal valence double ionization
potential values of He and alkaline earth metal (Be, Mg, Ca, Sr and Ba) atoms. Our
results are compared with the results available from the national institute of
standards and technology (NIST) database and other $\it {ab\, initio}$ calculations.
We have achieved an accuracy of $\sim0.1\%$, which is an improvement over the first
principles T-matrix calculations [J. Chem. Phys. {\bf 123}, 144112 (2005)]. We also
present results using the second-order many-body perturbation theory and the 
random-phase approximation in the equation-of-motion framework and these results are 
compared with the DI-EOMCC results.
\end{abstract}


\maketitle

Recent advances in the experimental techniques, such as x-ray free electron laser of
Linac Coherent Light sources of SLAC \cite{1,3} and attosecond pulses \cite{4,5},
have enabled studies of multi-ionization processes. The double photo-ionization of
atoms in which two electrons are ejected to continuum orbitals is a three-body quantal
problem. The complex interplay between the relativistic effects and the electron correlation
is of central importance in the accurate description of these processes \cite{6}. The advent 
of sub-femtosecond laser technology with the generation of attosecond pulses from the vacuum 
ultraviolet to the extreme ultraviolet wavelength region has opened up new 
perspectives on the observation of the correlated electron dynamics involved in the studies 
of the double ionization processes \cite{chen, krausz}. One of the outstanding theoretical
problems in these studies is to explain the simultaneous double ionization mechanisms 
 \cite{raphael}, which are different than the sequential ionization events. 
There has been experimental progress in the direction of attosecond tracing of the correlated electron-emissions 
in the non-sequential double ionization processes in atomic systems which requires a suitable 
theory that could describe the effects of the dynamical electron corrections adequately \cite{bergues}. To complement 
the sophisticated experimental techniques, it is desirable to have accurate theoretical methods to 
treat the double ionization continua. Attempts have been made using the T-matrix \cite{7,8}, delta 
self-consistent-field \cite{8a}, and with the four-component two-particle propagator methods \cite{10,11}. 
It is well known that not only the electron correlation, but also relaxation effects plays significant 
role in the accurate description of atomic states. Therefore measurements and 
calculations based on the lower-order  
many-body methods do not agree with each other \cite{seakins}. The equation-of-motion coupled-cluster 
(EOMCC) method \cite{12,13,14} provides a balanced treatment of the electron correlation and relaxation 
effects to determine the atomic states and also calculating differences of the energies in a direct 
manner. It uses a large configurational space constructed from the occupied and virtual spinors 
that takes into account both the static and dynamic correlations simultaneously. The wave functions and
energies of all the states of interest are obtained through the diagonalization of a similarity
transformed Hamiltonian, which in a sense, associated with the multi-reference theory but the EOMCC
is operationally single reference in nature. The (0,1) sector
Fock space multi-reference (FSMRCC) \cite{15,16} theory is equivalent to EOMCC method for the single 
ionization problem (EOMCC-IP) \cite{16a}, but this is not the case for the (0,2) sector FSMRCC 
and the DI-EOMCC methods. The FSMRCC theory uses the amplitude equations of all the lower sectors along 
with the amplitudes of that particular sector, whereas EOMCC requires the amplitudes of the sector of 
interest and those of the (0,0) sector.

The electronically excited molecules or atoms relax through various radiationless 
mechanisms by emitting electrons. The precise values of ionization potentials (IPs) and double
ionization potentials (DIPs) are important to analyze these relaxation processes and to
understand the excited states of the emitter \cite{17}. Furthermore, these processes play 
significant roles in designing efficient radio oncology schemes, which can be used as the
powerful tools to investigate the genotoxic effects on the living tissues \cite{18}.
Bartlett and coworkers were the first to implement the EOMCC method to calculate DIPs \cite{18a}, 
however their work was in the non-relativistic regime. Recently, we have implemented the relativistic 
EOMCC method and used it to calculate IPs of various atomic systems \cite{20}. In this work,
we extend the idea of single electron ionization using the relativistic EOMCC method to the domain of 
the double ionization spectra by considering simultaneous removal of two electrons from the closed-shell 
atomic systems.

\begin{figure}[t]
\includegraphics[width=7.5cm, height=1.25cm]{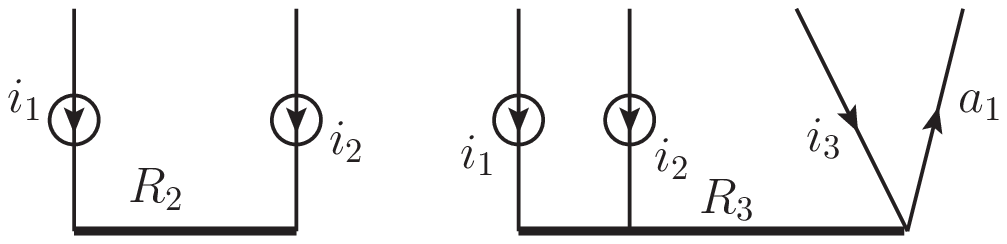}
\caption{Diagrammatic form of the $R_{2}$ and $R_{3}$ operators. Lines with down and up arrows denote 
for the occupied and unoccupied orbitals, respectively. Arrow with circle means a detached occupied orbital.}  
\label{fig2}
\centering
\begin{center}
\includegraphics[width=8.5cm, height=3.5cm]{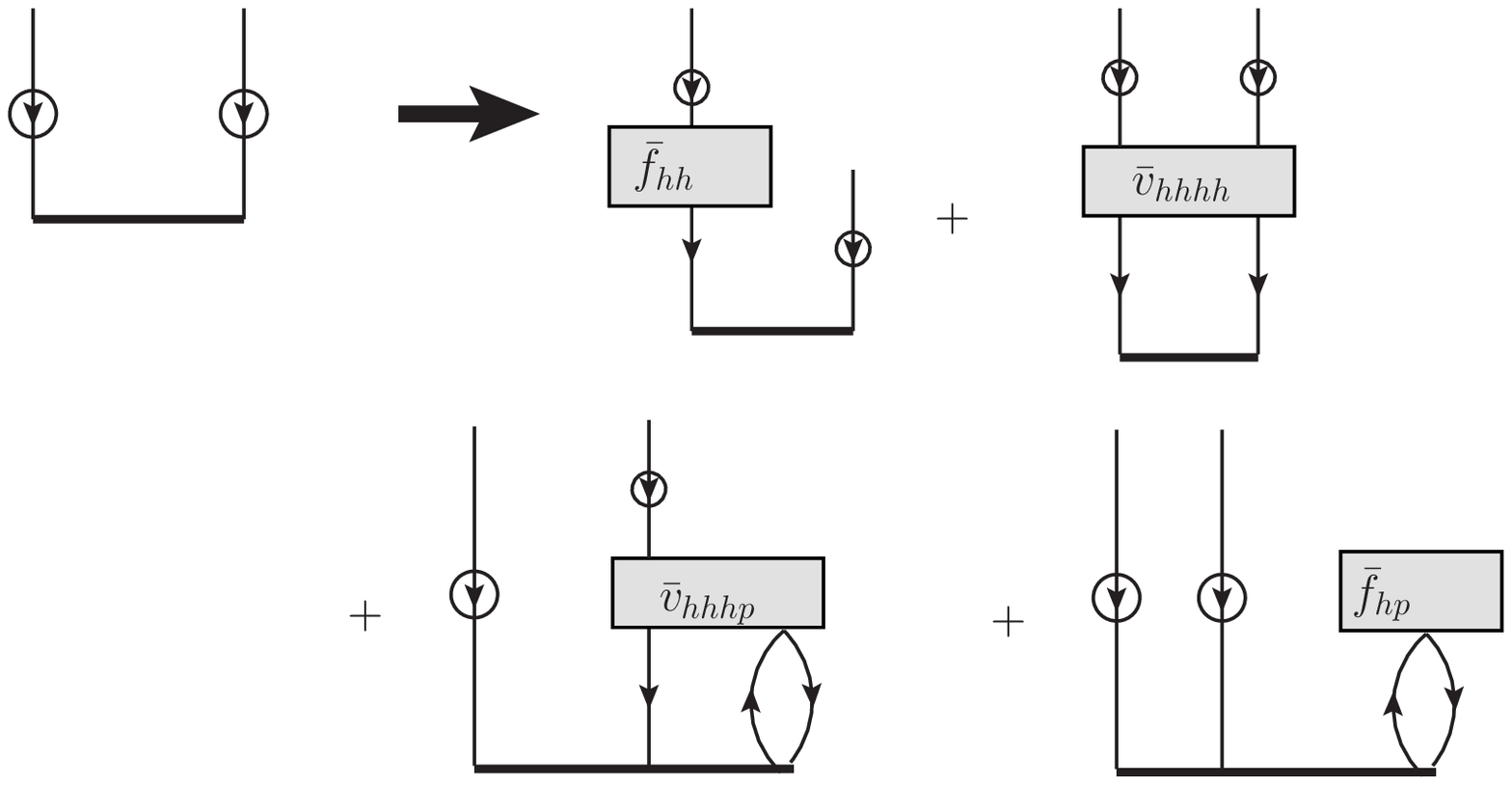}
\end{center}
\caption{Diagrams contributing to the 2h block.}  
\label{fig3}
\end{figure}

\begin{figure}[t]
\centering
\begin{center}
\includegraphics[width=8.5cm, height=4.5cm]{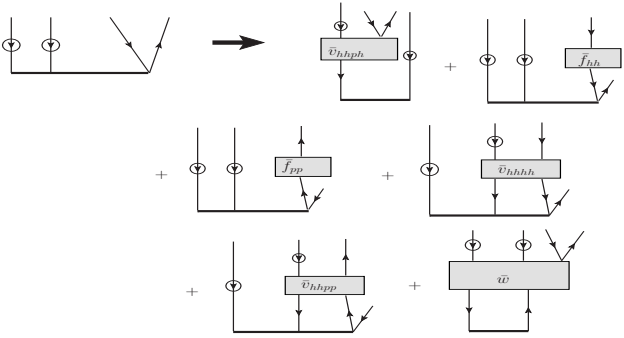}
\end{center}
\caption {Diagrams contributing to the 3h-1p block.} 
\label{fig4}
\end{figure}
 To the best of our knowledge, no
prior numerical results are available for the double ionization spectra of the atomic systems
using any variant of the relativistic coupled cluster theory. Development of the relativistic
DI-EOMCC method is a step forward as it can be used for studying various photo-ionization
spectra, highly energetic electron scattering processes and various electronic decay processes,
specially the Auger spectra in the atomic systems. The alkaline earth-metal atoms
are well suited for studying double ionization spectra, as the outer valence electrons are
well separated from the rest of the electrons. The He atom is also similar 
to these atoms, though they have different radial structures of the $ns$ orbitals. 
As a first application of our newly implemented 
relativistic DI-EOMCC method, we have calculated the valence DIPs of He, Be, Mg, Ca, Sr and Ba atoms. 
The computed results are compared with the values from the  
NIST database \cite{21} and with the available calculations based on the first principles T-matrix 
approach \cite{8}. We would also like to mention that, though we have calculated 
valence DIPs of the closed-shell atoms, the implemented DI-EOMCC method is applicable to both 
closed-shell and open-shell atomic and molecular systems having any number of valence electrons.

\begin{table*}[t]
\caption{The number and the $\alpha_0$ and $\beta$ parameters used for the GTOs to generate single particle orbitals 
at the DF level.}
\begin{ruledtabular}
\begin{tabular}{cccccccccccc}
Atom&Number of orbitals&\multicolumn{2}{c}{$s$}&\multicolumn{2}{c}{$p$}& 
\multicolumn{2}{c}{$d$}&\multicolumn{2}{c}{$f$}&\multicolumn{2}{c}{$g$}\\
&&$\alpha_{0}$&$\beta$&$\alpha_{0}$&$\beta$  
&$\alpha_{0}$&$\beta$&$\alpha_{0}$&$\beta$&$\alpha_{0}$&$\beta$\\
\cline{3-4} \cline{5-6} \cline{7-8} \cline{9-10} \cline{11-12} \\
\hline
He&\,(36s,35p,34d,33f,32g)&\,0.00075&\,2.075 &\,0.00155&\,2.080 &\,0.00258&\,2.180&\,0.00560&\,2.300&\,0.00765&\,2.450\\
Be&\,(36s,35p,34d,33f,32g)&\,0.00500&\,2.500&\,0.00615&\,2.650&\,0.00505&\,2.550&\,0.00500&\,2.530&\,0.00480&\,2.500 \\
Mg&\,(35s,34p,33d,32f,31g)&\,0.02950&\,1.630&\,0.09750&\,1.815&\,0.00750&\,2.710&\,0.00780&\,2.730&\,0.00800&\,2.750\\
Ca&\,(35s,34p,33d,32f,31g)&\,0.00895&\,2.110&\,0.00815&\,2.150&\,0.00750&\,2.500&\,0.00700&\,2.550&\,0.00690 &\,2.600\\
Sr&\,(35s,34p,33d,32f,30g)&\,0.01850&\,2.030&\,0.04750&\,2.070&\,0.00910&\,2.090&\,0.00950&\,2.100&\,0.00950&\,2.300\\
Ba&\,(35s,34p,33d,32f,31g)&\,0.00925&\,2.110&\,0.00975&\,2.040&\,0.00995&\,2.010&\,0.01015&2.035&\,0.01035&\,2.038\\
\end{tabular}
\end{ruledtabular}
\label{tab1}
\end{table*}
We also present results using two intermediate schemes at the second-order many-body perturbation
theory (MBPT(2)) and the random-phase approximation (RPA) level in the EOMCC framework to assess the
roles of the electron correlation effects. The 
former uses first order perturbed wave function, which corresponds to MBPT(2) energy as the ground 
state energy. For the latter, the effective Hamiltonian matrix elements are constructed only in the 
two hole ($2h$) space. It is clear that in both these approaches, the electron correlation effects 
are not treated as comprehensively as they are in the four-component all electron DI-EOMCC approach.

The starting point for the EOMCC method is the single reference CC wave function ${|\Psi_{0}\rangle}$, 
which is of the form 
\begin{equation}
{|\Psi_{0}\rangle=e^{\hat T_{n}}|\Phi_{0}\rangle},
\end{equation}
where ${|\Phi_{0}\rangle}$ is the Dirac-Hartree-Fock (DF) reference determinant and the cluster
operators are of the form
\begin{equation}
 \;\;\;\; {\hat T_n  = 
\sum\limits_{\stackrel{a_1 < a_2 \dots < a_n}{i_1 < i_2 \dots < i_n}}^M
t^{a_1 a_2 \dots a_n}_{i_1 i_2 \dots i_n}
 \ a^+_1 i_1 a^+_2 i_2 \dots a^+_n i_n }\;\;\;,
\end{equation}
with $i,\dots(a,\dots)$ corresponds to the strings of creation and annihilation operators acting on the reference determinant of
$M$ numbers of occupied electrons and stands for the hole and particle indices respectively. Projection onto the excited determinants 
${| \Phi^{a_1 a_2 \dots a_n}_{i_1 i_2 \dots i_n} \rangle = a^+_1 i_1 a^+_2 i_2 \dots a^+_n i_n | \Phi_{0} \rangle}$,
we get the simultaneous nonlinear algebraic equations for the correlation energy, defined as $ \Delta E_{corr}= E_g - E_{DF}$
for the ground state energy $E_g$ of the state $|\Psi_0 \rangle$ and DF energy $E_{DF}$ of the state $|\Phi_0\rangle$,
and also for the unknown cluster amplitudes of any order of excitations
\begin{equation}
{\langle \Phi^{a_1 a_2 \dots a_n}_{i_1 i_2 \dots i_n} | 
 (\hat{H}_N e^{\hat{T}})_c | \Phi_0 \rangle = \Delta E_{corr} \delta_{n,0}, \;\;\;\; (n=0, \cdots k) }.
\end{equation}

\begin{table}[t]
\caption{DF (${E_{DF}^{0}}$) and correlation energies from the MBPT(2) (${\Delta E_{corr}^{(2)}}$) and
CCSD (${\Delta E_{corr}^{(ccsd)}}$) methods along with the number of active orbitals used.}
\begin{ruledtabular}
\begin{tabular}{lrrrrrrrrrrrr}
Atom & \multicolumn{5}{c}{No. of active orbitals} & {${E_{DF}^{0}}$} & 
{${\Delta E_{corr}^{(2)}}$}& {${ \Delta E_{corr}^{(ccsd)}}$}  \\
 \cline{2-6} \\
& ${s}$  & ${p}$ & ${d}$ & ${f}$  
& ${g}$   \\
\hline
& & \\
He   &\,17   &\, 15   &\, 13  &\, 9  &\, 7  &\,   -2.8618  &\,  -0.0365  &\, -0.0416\\
Be   &\,14   &\, 12   &\, 12  &\, 10  &\,10  &\,   -14.5758  &\,  -0.0748  &\, -0.0929\\
Mg   &\,20   &\, 14   &\, 12  &\, 11  &\,10  &\,  -199.9350  &\,  -0.4097  &\, -0.4195\\
Ca   &\,16   &\, 15   &\, 12  &\, 11 &\, 10  &\, -679.7100  &\,  -0.7515  &\, -0.7648\\
Sr   &\,16   &\, 13   &\, 13  &\, 12 &\, 10  &\, -3178.0797  &\,  -1.6530  &\, -1.5922\\
Ba   &\,16   &\, 15   &\, 14  &\, 12 &\,9    &\,  -8135.6428 &\,  -2.2556   &\,-2.1258\\
\end{tabular}
\end{ruledtabular}
\label{tab2}
\end{table}
In the above equation, subscript $c$ means connected, $n$ is the level of excitations from
the DF state and $\hat{H}_N= \hat H- \langle \Phi_0 | \hat H | \Phi_0 \rangle$ is the normal 
ordered form of the Dirac-Coulomb (DC) Hamiltonian ($\hat H$) which is given by
\begin{eqnarray}
\hat H &=& \sum_i \left [ c\mbox{\boldmath$\alpha$}_i\cdot \textbf{p}_i+(\beta_i -1)c^2 + V_{nuc}(r_i) +
\sum_{j>i} \frac{1}{r_{ij}} \right], \ \ \
\end{eqnarray}
where $\mbox{\boldmath$\alpha$}_i$ and $\beta_i$ are the usual Dirac matrices, $V_{nuc}(r_i)$ is
the nuclear potential and $\frac{1}{r_{ij}}$ is the electron-electron repulsion potential.
The single particle energies are evaluated with respect to the rest mass energy $(-c^2)$ of the electron.
Note that unless stated otherwise, we use atomic unit (au) in this paper. In our calculations, we only 
consider singles and doubles excitations ($n=2$) which is referred to as the CCSD method.  

In the DI-EOMCC approach, the wave function for the $\mu^{th}$ state of the doubly ionized system ($M-2$ electron) can be written as 
\begin{equation}
{|\Psi_{\mu} \rangle =R^{M-2}_{\mu}|\Psi_{0}\rangle}, \ \ \ \ \mu =1,2,\cdots .
\end{equation}
The $R^{M-2}_{\mu}$ is a linear operator and it is of the form 
\begin{equation}
\begin{split}
R^{M-2}_{\mu} =& R_{2}+R_{3}+ \cdots\\
             =&  \sum_{i_1<i_2} r_{i_1 i_2} i_2 i_1 +  \sum\limits_{\stackrel{a_1}{i_1 < i_2 < i_3}} r_{i_1 i_2 i_3}^{a_1} a_1^{+} i_3 i_2 i_1 + \cdots .
\end{split}
\end{equation}
In our approximation, we considered upto the $R_{2}$ and $R_{3}$ operators which are diagrammatically shown in Fig. \ref{fig2}.

The difference ($\Delta E_{\mu} = E_{\mu} -E_g$) between the energies of the ground and the doubly ionized states ($E_{\mu}$) are calculated
\begin{equation}
[\bar{H}_N,\hat R^{M-2}_{\mu}]|\Phi_{0} \rangle=\Delta E_{\mu} \hat R^{M-2}_{\mu}|\Phi_{0} \rangle \ \ \ \ \forall \mu, 
\end{equation}
on projecting onto the basis of excited determinants ($|\phi_{i_1i_2}\rangle$ and$|\phi_{i_1i_2i_3}^{a_1}\rangle$)
with respect to $|\phi_0\rangle$, yields the matrix form,

\begin{equation}
\bar{H}_N R=R\Delta E.
\end{equation} 
Here, $\bar{H}_N = (\hat H_{N} e^{T})_c$ is the similarity transformed effective Hamiltonian. In our present
implementation of DI-EOMCC method, the matrix elements of the ${\bar{H}_N}$ Hamiltonian are constructed in the 
$2h$ and $3h-1p$ space and diagonalized to obtain the desired eigenvectors and eigenvalues. The 
anti-symmetrized diagrams contributing to the $2h$ and $3h-1p$ blocks are presented in Figs. \ref{fig3} and 
\ref{fig4}, respectively. The diagrams corresponding to the $2h$ block is responsible for the RPA 
calculations. These diagrams are evaluated with the help of the one-body, two-body, and three-body 
intermediate diagrams. The evaluation of the intermediate diagrams requires the solution of the CCSD equations. 
The intermediate diagrams are constructed by contracting the one-body and two-body parts of the effective 
Hamiltonian with the converged ${T_{1}}$ and ${T_{2}}$ amplitudes. The one-body intermediate diagrams are inserted
in Figs. \ref{fig3} and \ref{fig4} as $\bar f_{hh}$, $\bar f_{pp}$, and $\bar f_{hp}$. There are four distinct 
types of the two-body intermediate diagrams required for the construction of the $\bar{H_{N}}$ matrix elements for 
the DI-EOMCC calculations. These are also used as $\bar v_{hhhh}$, $\bar v_{hhhp}$, $\bar v_{hhph}$ and
$\bar v_{hhpp}$ in the above two figures. The three-body intermediate diagram is also shown in Fig. \ref{fig4} as $\bar w$.

The dimension of the ${\bar H_{N}}$ matrix is very large ($nh^2+nh^3np$, $nh^2+nh^3np$) and it is not amenable to a 
full diagonalization scheme. Here $nh$ and $np$ denote the number of holes and particles, respectively. We have, 
therefore, used the Davidson algorithm \cite{22} which is an iterative scheme to diagonalize the ${\bar{H}_N}$ matrix.
It avoids computation, storage and diagonalization of the full matrix. The Davidson algorithm 
performs reasonably well for the whole spectrum, but we have solved only for the root
corresponding to the lowest eigenvalue.

\begin{table}[t]
\caption{Double-ionization energies of the alkaline earth-metal (Be, Mg, Ca, Sr, Ba) and He atoms in eV.}
\begin{ruledtabular}
\begin{tabular}{l c c c c c  }
 Atom     &   NIST            &   \multicolumn{3}{c}{This work} & T-matrix \\
\cline{3-5}\\
  &  \cite{21} & MBPT(2)&  RPA & EOMCC & \cite{8} \\
\hline
He & 79.0051 & 78.8596 & 79.0061 & 79.0092 &  \\

Be & 27.5338 & 27.0949 & 27.5595 & 27.5421 & 27.79 \\

Mg & 22.6815 & 22.4369 & 22.7655 & 22.6624 & 22.97  \\

Ca & 17.9848 & 17.9007 & 18.0893 & 17.9992 & 17.82  \\

Sr & 16.7251 & 16.7119 & 16.8502 & 16.7370 &  \\
Ba & 15.2154 & 15.2950 & 15.3546 & 15.2356 &  \\
\end{tabular}
\end{ruledtabular}
\label{tab3}
\end{table}
The orbital basis sets were generated using the proper kinetic balance between the large and small components of the 
wave function, where the large component of the wave function is constructed using the linear combinations of 
the Gaussian-type of orbitals (GTOs) as
{$g_{kp}^{L}(r)=C_{ki}^{L}r^{n_\kappa}e^{-\alpha_{p}r^{2}}$,}
 with the imposed condition 
{$\alpha_{p}=\alpha_{0}\beta^{p-1}$,}
where $p=0,1.....k$ with $k$ number of GTOs. The corresponding $\alpha_{0}$ and  $\beta$ parameters for all the atoms 
along with the number of GTOs generated at the DF level are given in Table \ref{tab1}. The orbital basis is 
constructed by taking into account primitive functions up to $g$-harmonics. It is impractical to use of all the 
orbitals generated in the DF method in a correlation calculation. Therefore, we have considered all the occupied 
orbitals and for the virtual orbitals a certain threshold energies, since the contributions from the high lying orbitals
is rather small owing to their large energy values.
 These orbitals  
are referred to as active orbitals. The two parameter finite-size Fermi charge density distribution nuclear model   
{$\rho_{nuc}(r)=\frac{\rho_{0}}{1+e^{(r-b)/a}}$,} is used to evaluate the nuclear potential,
here ${\rho_{0}}$ is the average nuclear density and $b$ is the half-charge radius and $a$ is related to
the skin thickness.
 
\begin{figure}[t]
\centering
\begin{center}
\includegraphics[height=4.0cm, width=.4\textwidth]{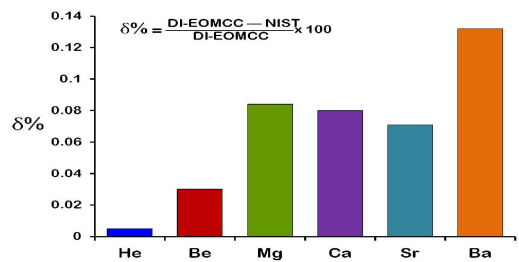}
\caption {(Color online) Relative deviations in ${\%}$ from the NIST values.}
\label{fig8}
\end{center}  
\end{figure}

In Table \ref{tab2}, we present the DF energy (${E^{0}_{DF}}$) and 
the correlation energies from the MBPT(2) (${E^{2}_{corr}}$) and the CCSD (${E^{CCSD}_{corr}}$)
methods along with the number of the active orbitals corresponding to each symmetry used in the calculations.
In Table \ref{tab3}, we present the results for the valence DIPs of He, Be, Mg, Ca, Sr, and Ba atoms. 
In fact, the formulation of the present method is valid even for determining DIPs of the states including 
inner occupied orbitals, however they require larger dimensional space and special care on accounting the
higher order correlation effects in the considered systems and left for the future work.  
All these results are compared with the tabulated values of NIST database \cite{21} and with the 
available results from the first principles T-matrix calculations \cite{8}. The differences between the DI-EOMCC (given in the table as EOMCC)
results and the NIST data are plotted as $\delta$ (in \%) in Fig. \ref{fig8}. All the calculated DIPs are in 
excellent agreement with those of the NIST data. The DI-EOMCC values differ by about 0.01 to 0.02 eV from the 
NIST values except for the He atom, where the agreement is even better. It is found that MBPT(2) underestimates
the values of DIPs relative to DI-EOMCC as the dominant part of the dynamic correlation is missing in this scheme. 
RPA, on the other hand, overestimates them except for the He atom. The 3h-1p block, which is the major source of the
non-dynamical correlation effects, is missing in the RPA scheme. Our DI-EOMCC results are more accurate than those 
obtained using the first principles T-matrix method. The results of the latter differ from the NIST values by about 
0.2 to 0.3 eV. 
\\
In conclusion, we have successfully implemented the relativistic EOMCC method to calculate the double ionization spectra.
As a first application, we have calculated the valence DIPs of the alkaline earth-metal
atoms along with the He atom. We have achieved an accuracy of $\sim0.1\%$, which are better than all previous DIP 
calculations. We have also compared the results of our MBPT(2) and RPA calculations with those of DI-EOMCC to
highlight the roles of the correlation effects to attain high-precision results.

H.P. and S.P. acknowledge a grant from CSIR XIIth Five Year Plan project on Multi-Scale Simulations of 
Material (MSM) and the resources of the Center of Excellence in Scientific Computing at CSIR-NCL. 


\end{document}